\documentclass[preprint,authoryear,12pt]{elsarticle}

 \usepackage{graphicx}
 \usepackage{booktabs}

\usepackage{amssymb}
\usepackage{setspace} 
\newtheorem{theorem}{Theorem}

\journal{Environmetrics}

\begin{document}

\begin{frontmatter}



\title{Probabilistic Forecasts of Solar Irradiance by Stochastic Differential Equations}


\author[address1]{Emil B. Iversen\corref{cor1}}
\ead{jebi@dtu.dk}
\author[address1]{Juan M. Morales}
\ead{jmmgo@dtu.dk}
\author[address1]{Jan K. M\o ller}
\ead{jkmo@dtu.dk}
\author[address1]{Henrik Madsen}
\ead{hmad@dtu.dk}

\address[address1]{Technical University of Denmark, Asmussens Alle, building 322, DK-2800 Lyngby, Denmark}
\cortext[cor1]{Corresponding author. Tel.: +45 60 67 19 85}

\begin{abstract}
Probabilistic forecasts of renewable energy production provide users with valuable information about the uncertainty associated with the expected generation. Current state-of-the-art forecasts for solar irradiance have focused on producing reliable \emph{point} forecasts. The additional information included in probabilistic forecasts may be paramount for decision makers to efficiently make use of this uncertain and variable generation. In this paper, a stochastic differential equation (SDE) framework for modeling the uncertainty associated with the solar irradiance point forecast is proposed. This modeling approach allows for characterizing both the interdependence structure of prediction errors of short-term solar irradiance and their predictive distribution. A series of different SDE models are fitted to a training set and subsequently evaluated on a one-year test set. The final model proposed is defined on a bounded and time-varying state space with zero probability almost surely of events outside this space.
\end{abstract}


\end{frontmatter}



\section{Introduction}

The operation of electric energy systems is today challenged by the increasing level of uncertainty in the electricity supply brought in by the larger and larger share of renewables in the generation mix. Decision-making, operational and planning problems in electricity markets can be characterized by time-varying and asymmetric costs. These asymmetric costs are caused by the need to continuously balance the electricity system to guarantee a reliable and secure supply of power. An understanding of the underlying uncertainty is, therefore, essential to satisfactorily manage the electricity system. This introduces the need for forecasts describing the entire variation of the renewable generation.

Solar irradiance is a source of renewable energy and, along with wind and hydro, is taking shape as a potential driver for a future free of fossil fuels. The worldwide installed capacity of photovoltaic energy systems has seen a rapid increase from 9.5 GW in 2007 to more than 100 GW by the end of 2012 (\cite{solcelleudviklingen2012}). The energy generation from solar irradiance is subject to weather conditions and, as such, it constitutes a variable and uncertain energy source.

Current state-of-the-art forecasts for solar energy have focused on point forecasts, that is, the most likely or the average outcome. Such point forecasts, however, do not adequately describe the uncertainty of the power production. This is recognized by the abundance of significant works on probabilistic forecasting for wind power, see for ex. \cite{Pinson2007Non} and \cite{Zhou2013Application}. 

In the literature, a variety of different approaches have been taken to provide reliable solar power forecasts. A review of some of these approaches is found in \cite{pedro2012assessment}, where the persistence is compared to time-series models such as ARIMA models and different neural network models. Artificial neural networks are also used in \cite{chen2011online}, in combination with a weather type classification, to provide point forecasts of PV production. In \cite{lorenz2009irradiance}, a forecast method that makes use of a clear sky model and numerical weather predictions is developed, also accounting for orientation and tilt of the PV panel. The paper by \cite{bhardwaj2013estimation} introduces a hidden Markov model for solar irradiance based on fuzzy logic. They exploit inputs such as humidity, temperature, air pressure and wind speed, among others. A time-series model for predicting one-hour-ahead solar power production is considered in \cite{yang2012hourly}. This paper employs a cloud cover index to model the absorption and refraction of the incoming light through the atmosphere. In \cite{bacher2009online}, an auto-regressive model with exogenous input is proposed. It predicts weighing the past observations and the numerical weather prediction and introduces a clear sky model to capture the diurnal variation.

Probabilistic forecasting of solar irradiance is, though, in its infancy. One work in this area is the one by \cite{Mathiesen2013Geostrophic}, where post-processing of numerical weather predictions are applied to obtain probabilistic forecasts. Previous work on stochastic differential equations and solar irradiance is, to the best of our knowledge, limited to \cite{Soubdhan2010Stochastic}, which formulates a very simple stochastic differential equation model for solar irradiance. As a consequence of its simplicity, the model was largely unsuccessful at forecasting. Stochastic differential equations are fruitfully used for wind power forecasting in \cite{Moller2013Probabilistic} by considering state-dependent diffusions and external input.

This paper describes a new approach to solar irradiance forecasting based on stochastic differential equations (SDEs). Modeling with SDEs has multiple benefits, among others:
\begin{itemize}
\item{SDE models are able to produce reliable point forecasts as well as probabilistic forecasts.}
\item{Model extensions are easy to formulate and have an intuitive interpretation. We can start with a simplistic model and extend it to a sufficient degree of complexity.}
\item{We can model processes that are bounded and assign zero probability to events outside the bounded interval, which is essential for correct probabilistic forecasts of solar irradiance.}
\item{We leave the discrete-time realm of Gaussian innovations and consider instead the more general class of continuous-time processes with continuous trajectories.}
\item{SDEs span a large class of stochastic processes with classical time-series models as special cases.}
\end{itemize}

The rest of this paper is structured as follows: Section 2 gives a general introduction to the stochastic differential equation framework and describes an estimation procedure. In Section 3, we discuss the proposed modeling approach. Section 4 starts with a simple SDE model to which new features are progressively added until a full-fledged model is obtained. In Section 5, the different models are compared to simple as well as complex benchmarks and the performance of the finished model is assessed. Lastly, Section 6 concludes the paper.

\section{Stochastic Differential Equations}

Suppose that we have the continuous time process $X_t \in \mathcal{X} \subset \mathbb{R}^n$. In general, it is only possible to observe continuous time processes in discrete time. We observe the process $X_t$ through an observation equation at discrete times. Denote the observation at time $t_k$ by $Y_k \in \mathcal{Y} \subset \mathbb{R}^l$ for $k \in \left\{ 0, \ldots, N \right\}$. Let the observation equation be given by:
\begin{eqnarray}
Y_k = h(X_{t_k}, t_k, e_k), \label{eq: observation equation}
\end{eqnarray}
where the variable $t_k$ allows for dependence on an external input at time $t_k$, $e_{k} \in \mathbb{R}^l$ is the random observation error, and $h(\cdot) \in \mathbb{R}^l$ is the function that links the process state to the observation. The simplest form of the observation equation is $h(\cdot) = X_{t_k} + e_k$.

\subsection{Definition of Stochastic Differential Equations}

In the ordinary differential equation setting, the evolution in time of the state variable $X_t$ is given by the deterministic system equation
\begin{eqnarray}
\frac{dX_t}{ dt} = f(X_t, t),
\end{eqnarray}
where $t \in \mathbb{R}$ and $f(\cdot) \in \mathbb{R}^n$. Complex systems such as weather systems are subject to random perturbations of the input or process that are not specified in the model description. This suggests introducing a stochastic component in the state evolution to capture such perturbations. This can be done by formulating the state evolution as a stochastic differential equation (SDE), as done in \cite{øksendal2010stochastic}. Thus, we can formulate the time evolution of the state of the process by the form:
\begin{eqnarray}
\frac{dX_t}{ dt} = f(X_t, t) + g(X_t, t) W_t, \label{eq: SDE initial form}
\end{eqnarray}
where $W_t \in \mathbb{R}^m$ is an $m$-dimensional standard Wiener process and $g(\cdot) \in \mathbb{R}^{n \times m}$ is a matrix function \citep{øksendal2010stochastic}. Multiplying with $dt$ on both sides of (\ref{eq: SDE initial form}) we get the standard SDE formulation:
\begin{eqnarray}
dX_t = f(X_t, t) dt + g(X_t, t) dW_t. \label{eq: SDE definition}
\end{eqnarray}
Notice here that we allow for a complex dependence on $t$, including external input at time $t$. While this form is the most common for SDEs, it is not well defined, as the derivative of $W_t$, $dW_t$, does not exist. Instead, it should be interpreted as an informal way of writing the integral equation:
\begin{eqnarray}
X_t = X_0 + \int_0^t{f(X_s, s) ds} + \int_0^t{g(X_s, s,) dW_s}. \label{eq: SDE definition integral form}
\end{eqnarray}
In Equation (\ref{eq: SDE definition integral form}), the behavior of the continuous time stochastic process $X_t$ is expressed as the sum of an initial stochastic variable, an ordinary Lebesgue integral, and an Ito integral.

In a deterministic ordinary differential equation setting, the solution would be a single point for each future time $t$. In the SDE setting, in contrast, the solution is the probability density of $X_t$ for any state, $x$, and any future time, $t$. For an It\={o} process given by the stochastic differential equation defined in (\ref{eq: SDE definition}) with drift $f(X_t,t)$ and diffusion coefficient $g(X_t,t)= \sqrt{2 D(X_t,t)}$, the probability density $j(x,t)$ in the state $x$ at time $t$ of the random variable $X_t$ is given as the solution to the partial differential equation known as the Fokker-Planck equation \citep{björk2009arbitrage}:
\begin{eqnarray}
    \frac{\partial}{\partial t}j(x,t) = -\frac{\partial}{\partial x}\left[f(x,t)j(x,t)\right] + \frac{\partial^2}{\partial x^2}\left[ D(x,t)j(x,t)\right]. \label{eg:Fokker-Plank}
\end{eqnarray}
Thus, given a specific SDE, we can find the density at any future time by solving a partial differential equation.

SDEs are a general class of processes. This is stated by the L\'{e}vy-It\={o} decomposition, which says that, under sufficient regularity conditions, all stochastic processes with continuous trajectories can be written as a SDE \citep{øksendal2010stochastic}. Hence many of the ordinary discrete-time stochastic processes can be seen as a SDE being sampled at discrete times, and therefore, SDEs is a generalization of generic time-series models in discrete time.

\subsection{Parameter Estimation}

In this section, we outline how to estimate parameters in a SDE of a general form and, in particular, with a state-dependent diffusion term. First, we go into detail on the estimation procedure of the parameters of a SDE with a state-independent diffusion term. Second, we show how to transform a process with state-dependent diffusion term into a process with a unit diffusion term, whereby the previously mentioned estimation procedure can be applied.

Consider the model defined by Equations (\ref{eq: observation equation}) and (\ref{eq: SDE definition}) given by:
\begin{eqnarray}
dX_t &=& f(X_t, t) dt + g(X_t, t) dW_t \label{eq: general SDE} \\
Y_k &=& h(X_{t_k}, t_k, e_k).
\end{eqnarray}
On the basis that we want to estimate the parameters in the above model, the problem can be formulated as follows: Find a parameter vector, $\hat{\theta} \in \Theta$, that maximizes some objective function of $\theta$. There are several possible choices for an objective function. A natural choice in this framework is to choose an objective function that maximizes the probability of seeing the observations given by $\mathcal{Y}_N = \left\{ Y_0, \ldots, Y_N \right\}$. This leads to choosing the likelihood function as objective function, i.e.,
\begin{eqnarray}
L \left( \theta ; \mathcal{Y}_N \right) = p \left( \mathcal{Y}_N | \theta \right) = \left( \prod_{k=1}^N{ p \left( Y_k | \mathcal{Y}_{k-1}, \theta \right) } \right) p(Y_0| \theta).
\end{eqnarray}
Even though this problem could, in principle, be solved using the Fokker-Planck equation, this is only feasible for systems with simple structures, as it involves solving a complex partial differential equation.

The estimation procedure, which we shall introduce next, relies on the system having a specific form, namely:
\begin{eqnarray}
dX_t &=& f(X_t, t) dt + g(t) dW_t \label{eq: CTSM system 1} \\
Y_k &=& h(X_{t_k}, t_k) + e_k. \label{eq: CTSM system 2}
\end{eqnarray}
In the system defined by Equations (\ref{eq: CTSM system 1}) and (\ref{eq: CTSM system 2}), we assume that $g(\cdot) \in \mathbb{R}^{n \times n}$ does not depend on the state $X_t$. Also, we assume that the observation noise is an additive Gaussian white noise, i.e., $e_k \sim \mathcal{N}\left( 0,S_k( t_k ) \right)$, where $S_k(t_k)$ is some covariance matrix, possibly depending on time. It is clear that restricting $g(\cdot)$ to not depend on $X_t$ limits our model framework severely. As we shall see, this can, to some degree, be remedied by a transformation using It\={o}-calculus. The restriction of having additive Gaussian measurement noise should be dealt with by transformations of the observations.

As the system defined by Equations (\ref{eq: CTSM system 1}) and (\ref{eq: CTSM system 2}) is driven by Wiener noise, which has Gaussian increments, and the observation noise is Gaussian, it is reasonable to assume that the density of $Y_{k}|\mathcal{Y}_{k-1}$ can be approximated by a Gaussian distribution. Note that the Gaussian distribution is completely characterized by its mean and covariance. This implies that using the extended Kalman filter, which is linear, is appropriate.

The one-step predictions for the mean and variance are defined as:
\begin{eqnarray}
\widehat{Y}_{k|k-1} &=& \mathbb{E}\left[Y_k | \mathcal{Y}_{k-1}, \theta \right]  \\
R_{k|k-1} &=& \mathbb{V} \left[Y_k | \mathcal{Y}_{k-1}, \theta \right],
\end{eqnarray}
where $\mathbb{E}\left[ \cdot \right] $ and $\mathbb{V} \left[ \cdot \right]$ denote the expectation and variance, respectively. The innovation is given by
\begin{eqnarray}
\epsilon_k = Y_k - \widehat{Y}_{k|k-1}.
\end{eqnarray}
Using this, we can now write the likelihood function as
\begin{eqnarray}
L \left( \theta ; \mathcal{Y}_N \right) = \left( \prod_{k=1}^N{ \frac{ \exp \left( - \frac{1}{2} \epsilon_k^{\top} R_{k|k-1}^{-1} \epsilon_k \right) }{\sqrt{ \det\left( R_{k|k-1} \right)} \left( \sqrt{2 \pi} \right)^l } } \right) p(Y_0| \theta),
\end{eqnarray}
where $l$ is the dimension of the sample space and $(\cdot)^{\top}$ denotes the vector transpose. The estimate of $\theta$ can be found by solving the optimization problem
\begin{eqnarray}
\hat{\theta} = \textrm{arg} \max_{\theta \in \Theta} \left( \log (L \left( \theta ; \mathcal{Y}_N \right) ) \right).
\end{eqnarray}

The Kalman gain governs how much the one-step prediction of the underlying state, $\widehat{X}_{k|k-1}$, should be adjusted to form the state update, $\widehat{X}_{k|k}$, from the new observation. This is given by
\begin{eqnarray}
K_k = P_{k|k-1} C^{\top} R_{k|k-1}^{-1},
\end{eqnarray}
where $C$ is the first order expansion of $h(\cdot)$, i.e., the Jacobian, and $P_{k|k-1}$ is the covariance of the one-step prediction. The state update is then given by
\begin{eqnarray}
\widehat{X}_{k|k} &=& \widehat{X}_{k|k-1} + K_k \epsilon_k  \\
P_{k|k} &=& P_{k|k-1} - K_k R_{k|k-1}^{-1} K_k^{\top}.
\end{eqnarray}
Hence, the state update is a combination of the previous state estimate and the new information obtained from the $k$'th observation, $Y_k$. 

The procedure of estimating the parameters has been implemented in the software tool described in \cite{Kristensen2004Parameter}.

\subsection{It\={o} Calculus and the Lamperti Transform}

We will now discuss how a SDE of the form in Equation (\ref{eq: general SDE}) can be transformed to the form in Equation (\ref{eq: CTSM system 1}) to allow for the estimation procedure previously introduced. The fundamental tool for the transformation of SDEs is It\={o}'s lemma, as stated in \cite{øksendal2010stochastic}. Below we introduce the 1-dimensional It\={o} formula and the Lamperti transform. The multidimensional It\={o} formula is covered in \cite{øksendal2010stochastic}. For a more detailed description of the Lamperti transform and how to apply it to multivariate processes, see \cite{Moller2010Lamperti}.

\begin{theorem}[The 1-dimensional It\={o} formula \label{thm: Ito formula}]
Let $X_t$ be an It\={o} process given by
\begin{eqnarray}
dX_t &=& f(X_t, t) dt + g(X_t, t) dW_t. \label{eq: 1-dim Ito}
\end{eqnarray}
Let $\psi(x,t) \in C^2( \left[ 0,\infty \right) ) \times \mathbb{R}$. Then
\begin{eqnarray}
Z_t &=& \psi(X_t,t)
\end{eqnarray}
is again an It\={o} process, and
\begin{eqnarray}
dZ_t &=& \frac{\partial \psi }{\partial t} (X_t,t) dt + \frac{\partial \psi }{\partial x} (X_t,t) dX_t + \frac{1}{2}\frac{\partial^2 \psi }{\partial x^2} (X_t,t) (dX_t)^2,
\end{eqnarray}
where $(dX_t)^2$ is calculated according to the rules
\begin{eqnarray}
dt \cdot dt = dt \cdot dW_t = dW_t \cdot dt = 0, \qquad dW_t \cdot dW_t = dt.
\end{eqnarray}
\end{theorem}

The It\={o} formula stated in Theorem \ref{thm: Ito formula} can be used to transform the process to a SDE with unit diffusion by the Lamperti transform.

\begin{theorem}[Lamperti transform]
Let $X_t$ be an It\={o} process defined as in (\ref{eq: 1-dim Ito}), and define
\begin{eqnarray}
\psi(X_t,t) &=& \left. \int{\frac{1}{g(x,t)}}dx \right|_{x=X_t}.
\end{eqnarray}
If $\psi$ represents a one to one mapping from the state space of $X_t$ onto $\mathbb{R}$ for every $t \in \left[ 0, \infty \right)$, then choose $Z_t = \psi(X_t,t)$. Then $Z_t$ is governed by the SDE
\begin{eqnarray}
dZ_t &=& \left( \psi_t(\psi^{-1}(Z_t,t),t) +  \frac{f(\psi^{-1}(Z_t,t),t)}{g(\psi^{-1}(Z_t,t),t)} \right. \\
 & & \left. - \frac{1}{2} g_x(\psi^{-1}(Z_t,t),t) \right) dt + dW_t,
\end{eqnarray}
where $g_x(\cdot)$ and $g_t(\cdot)$ denote the derivatives of $g(\cdot)$ with regard to $x$ and $t$, respectively, and $\psi_t$ denotes the derivative of $\psi$ with respect to $t$. 
\end{theorem}
This result is obtained by applying the It\={o} formula.

\section{Modeling Methodology}

In Section 2, we explained how to estimate the parameters of an SDE model, once we have a candidate model. In this section, we will consider how to arrive at a candidate model and how to extend it to overcome the deficiencies that may be identified. The modeling methodology consists of the following steps: data preprocessing, identifying model extensions, drawing inference between different models, and lastly, validation of the proposed model. This procedure is iterative and should be continued until the model passes the validation stage.

\subsection{Data Preprocessing}

Prior to developing the SDE model itself, it is essential to consider if the observation noise in Equation (\ref{eq: CTSM system 2}) is Gaussian. If this is not the case, the observations should be transformed so that this assumption holds. Furthermore, outlier detection and data aggregation should be done at this preliminary stage, as these considerations will affect the subsequent steps.

\subsection{Identifying Model Extensions}

The modeling is done by starting with a simple model and then identifying extensions to this simple model. The challenge is to find significant extensions without overparametrizing the model. We may infer model extensions in different ways. One way is to have a mechanistic understanding of the system, for example, by noticing that the states can not take on negative values. A second way is to identify model deficiencies and extend the model to overcome these deficiencies. One initial method of identifying model deficiencies is to consider the autocorrelation of the residuals. If there is significant autocorrelation, it indicates that not all information is captured by the model and the model should be extended accordingly. Another approach is to consider the size of the diffusion term, as large diffusion coefficients indicate model deficiencies in the corresponding state. Different simulation approaches may also help identifying extensions, for instance, looking at the long-term behavior of a simulated model may serve to highlight any deficiencies in the modeling.

Once a model is deemed deficient, one should ponder what can be done to remedy this deficiency. One approach is to consider one or more parameters of the model as random states. For example, we can introduce
\begin{eqnarray}
d \theta_{i,t} = \alpha_{\theta_{i}}(\mu_{\theta_{i}}-\theta_{i,t})dt +  \sigma_{\theta_{i}} dW_{\theta_{i}}
\end{eqnarray}
to describe the evolution of the parameter $\theta_{i,t}$. Thus, we allow this parameter to be centered around $\mu_{\theta_{i}}$, it may, however also deviate from this value. The model is then to be re-estimated with the introduced random state. This new model should then be tested for significantly improving the fit.

\subsection{Statistical Inference}

In this step, a candidate model is formulated and statistical testing is performed by comparing likelihoods, information criteria and statistics. Likelihood ratio tests are performed where applicable and are, as such, the preferred test. If two models are not nested, that is, one is not a sub-model of the other, information criteria such as AIC and BIC are used as well. We also consider model reductions in this step and remove non-significant parameters. It is, however, important to consider if non-significant parameters are needed to fulfill basic model requirements. These requirements could be technical constraints such as non-negative states or parameter values that lead to stable solutions. In these cases, it can be preferable to maintain insignificant parameters in the model.

\subsection{Evaluation}

In this stage, the developed model is evaluated against its objective. As we aim at probabilistic forecasting, the evaluation of the model should focus on this, and therefore, should differ from the evaluation of point forecast models. Since the objective is to capture the predictive density, we compare its quantiles with the empirical coverage rates. As part of the evaluation stage, we may again look for model deficiencies and extensions. If there are obvious extensions or deficiencies, we may conclude that the modeling is not finished. We may want to forecast for different horizons and to consider the prediction intervals. Visual inspections of the predictions may also provide useful insight into the performance of the model. Simulating the long-term behavior of a fitted model can also provide useful information. More specifically, simulation techniques such as Markov chain Monte Carlo can give an insight into the performance of the model, for example, considering simulated stationary distribution versus empirical.


\section{Solar Irradiance}

In this section, we apply the theory from Section 2 and the methodology from Section 3 to model solar irradiance. We aim at obtaining a model that correctly describes its predictive density.

\subsection{Data}

The data set at our disposal belongs to a meteorological station located in the western part of Denmark. The data include hourly observations of irradiance on a flat surface together with predictions for irradiance based on a numerical weather prediction model from the Danish Meteorological Institute. The numerical weather prediction (NWP) provides a 48-hour forecast of the irradiance, which is updated every 6 hours. We use the most recent forecast in the model. The data covers a period of three years from 01/01-2009 to 31/12-2011. We divide the period into a training and a test set, with the training set covering the first two years and the test set the last year.

\subsection{Model 1: Tracking the NWP}

We start by introducing a simple SDE model for solar irradiance that tracks the numerical weather prediction (NWP) provided by the Danish Meteorological Institute, i.e.,
\begin{eqnarray}
dX_t &=& \theta( \textrm{NWP}_t \mu - X_t) dt + \sigma_x dW_t \label{eq: simple SDE 1} \\
Y_k &=& X_{t_k} +  \epsilon_k \label{eq: simple SDE 2}.
\end{eqnarray}
In this model and the following, we denote the observed solar irradiance at time $t_k$ by $Y_k$. $\textrm{NWP}_t$, is an external input representing the predicted irradiance at time $t$. In the model, we have parameter $\mu$, which allows for a local scaling of the $ \textrm{NWP}_t$, such that it does not over or under shoot on average. The parameter $\theta$ determines how rapidly the model reverts to the predicted level of irradiance. The system noise is controlled by parameter $\sigma_x$. The observation error is denoted $\epsilon_k$ and is a stochastic variable with distribution $\mathcal{N}(0,\sigma_{\epsilon})$.


\subsection{Model 2: Scaling with Maximum Irradiance}

Solar irradiance is highly cyclical. To capture this, we extend the simple SDE model defined in (\ref{eq: simple SDE 1})-(\ref{eq: simple SDE 2}) by introducing the maximum irradiance in hour $t$, $\textrm{Max}_t$, as a scaling factor. This leads to a formulation where we let the stochastic process $X_t$ denote the proportion of  extra terrestrial irradiance (i.e., the irradiance that would arrive at the surface if there were no atmosphere) that reaches the surface. Thus, we formulate the following SDE model:
\begin{eqnarray}
dX_t &=& \theta \left( \frac{\textrm{NWP}_t}{\textrm{Max}_t} \mu - X_t \right) dt + \sigma_x dW_t \\
Y_k &=& \textrm{Max}_{t_k} X_{t_k} +  \epsilon_k 
\end{eqnarray}
The above process is, however, undefined at night, when $\textrm{Max}_t = 0$. To overcome this, we can instead think of $X_t$ as a process that describes the state of the atmosphere and how much solar irradiance there would potentially be allowed through. In this context, it clearly makes sense to have $X_t$ defined at night. Thus, we can solve the issue of having $\textrm{Max}_t = 0$ by adding a small constant, say $\delta = 0.01$. Given that the $\textrm{NWP}_t$ is also equal to zero at night, we introduce another parameter $\beta$ (to be estimated) that is added to $\textrm{NWP}_t$ such that $X_t$ is not forced to tend to zero at night. This leads to the model:
\begin{eqnarray}
dX_t &=& \theta \left( \frac{\textrm{NWP}_t + \beta }{\textrm{Max}_t + \delta } \mu - X_t \right) dt + \sigma_x dW_t \\
Y_k &=& \textrm{Max}_{t_k} X_{t_k} +  \epsilon_k,
\end{eqnarray}
where there is no issue of dividing by zero.

\subsection{Model 3: Introducing a Lower Bound}

Based on our previous consideration regarding Model 2, it is apparent that we must have $X_t \in \left[ 0,1 \right]$. Enforcing a lower bound on the process is easily done by introducing a state-dependent diffusion, where the diffusion term decreases to zero as the process approaches the bound. One such model could be the following:
\begin{eqnarray}
dX_t &=& \theta \left( \frac{\textrm{NWP}_t + \beta }{\textrm{Max}_t + \delta } \mu - X_t \right) dt + \sigma_x X_t dW_t \\
Y_k &=& \textrm{Max}_{t_k} X_{t_k} +  \epsilon_k. 
\end{eqnarray}
Note that the diffusion term becomes small as the process gets closer to zero. As a result, the drift term dominates the process under these circumstances. Furthermore, since we assume that $\theta > 0$ and have that $ \frac{\textrm{NWP}_t + \beta }{\textrm{Max}_t + \delta }  > 0$, the drift term eventually pulls the process away from zero in such a case.

In the estimation procedure we have assumed that the noise is non-state dependent, which is clearly not the case here. Therefore, we need to work with the Lamperti transformed process. The Lamperti transformation is given by:
\begin{eqnarray}
Z_t &=& \psi \left(X_t,t \right) = \left. \int{\frac{1}{\sigma_x x}}dx \right|_{x=X_t} = \frac{\log{ \left(X_t \right)}}{\sigma_x} \\
X_t &=& \psi^{-1} \left(Z_t,t \right) = e^{\sigma_x Z_t}.
\end{eqnarray}
Noting that $\psi_t(\cdot) = 0$ and $g_x(\cdot) = \sigma_x$, we can now make use of the Lamperti transform to obtain the process on the transformed $Z$-space, which becomes:
\begin{eqnarray}
dZ_t &=& \left( \frac{ \theta \left( \frac{\textrm{NWP}_t + \beta }{\textrm{Max}_t + \delta } \mu - e^{\sigma_x Z_t} \right)}{\sigma_x e^{\sigma_x Z_t} } - \frac{\sigma_x}{2}  \right) dt + dW_t \\
Y_k &=& \textrm{Max}_{t_k} e^{\sigma_x Z_{t_k}}+  \epsilon_k.
\end{eqnarray}

In the sequel, we shall only state the model in the original domain and not in the Lamperti transformed domain, as they are equivalent in the sense of yielding the same output.

\subsection{Model 4: Introducing an Upper Bound}

As for the lower bound, we can introduce an upper bound, where the diffusion term decreases to zero as the process approaches the bound. One such model could be the following:
\begin{eqnarray}
dX_t &=& \theta \left( \frac{\textrm{NWP}_t + \beta }{\textrm{Max}_t + \delta } \mu - X_t \right) dt + \sigma_x X_t ( 1 - X_t) dW_t \\
Y_k &=& \textrm{Max}_{t_k} X_{t_k} +  \epsilon_k .
\end{eqnarray}
where $\theta > 0$ and $\frac{\textrm{NWP}_t + \beta }{\textrm{Max}_t + \delta }  < 1$. Therefore, the process will be moved away from the upper bound as the diffusion term vanishes when the process approaches one.

Other diffusion terms can also be implemented to include noise structures. Models such as $g(\cdot) = \sigma_x \sqrt{x(1-x)}$, $g(\cdot) = \sigma_x (1-x)  \sqrt{x}$ or $g(\cdot) = \sigma_x x \sqrt{1 - x}$ can capture diffusion terms with different degrees of peakedness and skewness. Also, more general forms can be found, these involving more often than not long and tedious calculations. From out of the models investigated here, the one with diffusion term $g(\cdot) = \sigma_x x ( 1-x) $ performs the best.

\subsection{Model 5: Scaling the Upper Bound}

In Models 2, 3 and 4 we have assumed the extra terrestrial irradiance, $\textrm{Max}_t$, as the upper limit for the solar irradiance. It should be clear, however, that this level can never be attained, because there will always be some refraction by the atmosphere. Hence, we can possibly scale down the upper limit, $\textrm{Max}_t$, to improve the model. This is done by introducing a factor, $\gamma$, on the maximum solar irradiance. In doing so, the model becomes:
\begin{eqnarray}
dX_t &=& \theta \left( \frac{\textrm{NWP}_t + \beta }{ \gamma \textrm{Max}_t + \delta } \mu - X_t \right) dt + \sigma_x X_t ( 1 - X_t) dW_t \\
Y_k &=&\gamma \textrm{Max}_{t_k} X_{t_k} +  \epsilon_k.
\end{eqnarray}

\subsection{Model 6: Introducing a Stochastic Time-Constant}

It might also be useful to let the time constant $\theta$ vary over time, which reflects that sometimes the numerical weather prediction performs well at predicting the irradiance and other times its performance is not as good. Furthermore, there is a lag in the numerical weather prediction as it takes 4  hours to solve the NWP model. Moreover, the NWP model is only run every 6 hours. This leads to the NWP being between 4 and 10 hours old. We take the exponential to the time-varying coefficient to avoid negative values of $\theta$ that would make the process unstable and diverge from the meteorological prediction. The resulting model is as follows:
\begin{eqnarray}
dX_t &=& e^{A_t} \left( \frac{\textrm{NWP}_t + \beta }{\gamma \textrm{Max}_t + \delta } \mu - X_t \right) dt + \sigma_x X_t ( 1 - X_t) dW_{1,t} \\
dA_t &=& \theta_A(\mu_A - A_t)dt + \sigma_A dW_{2,t} \\ 
Y_k &=& \gamma \textrm{Max}_{t_k} X_{t_k} +  \epsilon_k .
\end{eqnarray}
We have introduced the stochastic process $A_t$, which reverts to the level $\mu_A$. The speed at which this reversion occurs is determined by $\theta_A$, while the system noise is governed by $\sigma_A$. Other parameters such as $\mu$ or $\beta$ could also be chosen to vary over time. This possibility has been considered as well. However, the selected model performs the best.

\subsection{Model 7: Introducing a Diurnal Variation}
There might be variations over the day in the accuracy of the numerical weather prediction. Accordingly, we introduce a term that captures this diurnal variation as follows:
\begin{eqnarray}
dX_t &=& e^{A_t} \left( \frac{\textrm{NWP}_t + \beta }{\gamma \textrm{Max}_t + \delta } \left( \mu - \omega_1 \sin \left( \frac{2 \pi}{24} t + \omega_2 \right) \right) - X_t \right) dt  \\
&& + \sigma_x X_t ( 1 - X_t) dW_{1,t} \nonumber \\
dA_t &=& \theta_A(\mu_A - A_t)dt + \sigma_A dW_{2,t} \\ 
Y_k &=& \gamma \textrm{Max}_{t_k} X_{t_k} +  \epsilon_k.
\end{eqnarray}
Firstly, note that we use a sinusoid to describe a periodic behaviour. Secondly, we work in hourly time steps, which explains $\frac{2 \pi}{24} t$. We then introduce a period shift, $\omega_2$, and an amplitude, $\omega_1$. The sinusoid is added to the scaling of the meteorological prediction, which translates into the $\textrm{NWP}_t$ being more accurate in some hours of the day than in others.

We have ended up with a model that includes a maximum hourly irradiance, a numerical weather prediction as external input, stochastic time constants, and a non-Gaussian system noise that confines the process between zero and the extra terrestrial irradiance. In the following section, validation results of the final model are presented.


\section{Model Validation}

In this section, the different models are fitted to the data pertaining to the training set and evaluated in terms of their likelihood and information criteria. For model validation, we consider the performance of the fitted models in terms of likelihood on the test set. As we are concerned with the conditional distribution of the irradiance at a future time, traditional point-forecasting metrics, such as mean absolute error (MAE) or root mean square error (RMSE) are not appropriate, because they consider only the deviation from the point forecast. Also, the volatility of the models changes over time and depending on the state of the process, which is not taken into account in MAE or RMSE. As the likelihood is a proper scoring rule (\cite{gneiting2007strictly}), meaning that a better fit of the data will result in a better score, the likelihood is used in the fitting procedure. We choose the likelihood over other proper scoring rules such as the continuous rank probability score, as the likelihood in our case is a direct result of the optimization algorithm.

To compare the models with more classical alternatives, we consider an autoregressive model with external input (ARX) and an autoregressive model with external input and time-varying system variability, where the volatility is modeled using a generalized linear model (ARX-GLM). The ARX model is specified as follows:
\begin{eqnarray}
Y_{k+1} = \theta_1 Y_{k} + \theta_2 \textrm{NWP}_{k+1} + \epsilon_{k+1}, \qquad \textrm{where } \epsilon_{k+1} \sim \mathcal{N}(0,\sigma^2).
\end{eqnarray}
The ARX-GLM model takes the form
\begin{eqnarray}
Y_{k+1} = \theta_1 Y_{k} + \theta_2 \textrm{NWP}_{k+1} + \tilde{\epsilon}_{k+1}, \qquad \textrm{where } \tilde{\epsilon}_{k+1} \sim \mathcal{N}\left(0,f_{\tilde{\epsilon}}(\cdot)^2\right).
\end{eqnarray}
We find, after a fitting procedure, that an appropriate form of the variance scaling in the generalized linear model is
\begin{eqnarray}
f_{\tilde{\epsilon}}(k+1) = \sigma \left( \textrm{Max}_{t_{k+1}} \right)^{3/4},
\end{eqnarray}
where $\textrm{Max}_{t_{k+1}}$ is specified as in Model 2. For a general introduction to generalized linear models see \cite{madsen2011introduction}.

Additionally, we benchmark against climatological forecasts. The na\"{\i}ve forecast method is to use the empirical distribution, with no time dependence, to predict the solar irradiance. A slightly less naive approach is to use the empirical distribution of irradiance as a function of hour-of-day. A third climatological benchmark is to use both hour-of-day and month-of-year to predict the distribution. These benchmarks are clearly na\"{\i}ve as they do not use the previous observation to predict. Also, as the climatological approach is non-parametric, we use the empirical likelihood to evaluate the fit.

\begin{table}[!ht]
\small
\begin{center}
    \begin{tabular}{rrrrrrrrr}
    \toprule
         & & \multicolumn{3}{c}{Training Set} & \multicolumn{1}{c}{Test Set}\\  \cmidrule(r){3-5} \cmidrule(r){6-6}
                        & d.f.  & LL 	    & AIC & BIC &  LL  \\ \midrule 
    Clim.1              & -   & -96397  & - & - &  -48421   \\                        
    Clim.2              & -   & -65060  & - & - &  -33801    \\
    Clim.3              & -   & -48038  & - & - &  -24547    \\ 
    ARX                 & -   & -95080  & - & - &  -47411    \\                
    ARX-GLM             & -   & -39869  & - & - &  -20042    \\     
    Model 1             & 5   & -95619  & 191248 & 191287 &  -46621    \\
    Model 2             & 6   & -31042  & 62096  & 62143  &  -15277    \\
    Model 3             & 6   & -31993  & 63998  & 64045  &  -15650    \\ 
    Model 4             & 6   & -30596  & 61204  & 61251  &  -14925    \\
    Model 5             & 7   & -30495  & 61004  & 61058  &  -14860    \\
    Model 6             & 10  & -30370  & 60760  & 60838  &  -14816    \\ 
    Model 7             & 12  & \textbf{-30267}  & \textbf{60554}  & \textbf{60625}  &  \textbf{-14719}    \\ \bottomrule
\end{tabular}
\end{center}
\vspace{-10pt}
  \caption{\emph{In this table the log-likelihood of the different models are shown on the training and test set along with information criteria and degrees of freedom. The climatological predictors are evaluated in terms of empirical likelihood.}}
\label{tab:ModelStats}
\end{table}

The results of the different models are presented in Table \ref{tab:ModelStats}. We see that Model 7 best describes the data in the training set, as well as in the test set. Furthermore note that the improvement from the quite na\"{\i}ve Model 1 to Model 2 is huge, which justifies the change in the state space. Notice also the correspondence between the ARX model and Model 1. Indeed, the ARX model is the similar to a discrete version of Model 1.

\begin{figure}[h!]
\begin{center}
  \includegraphics[width=1\textwidth]{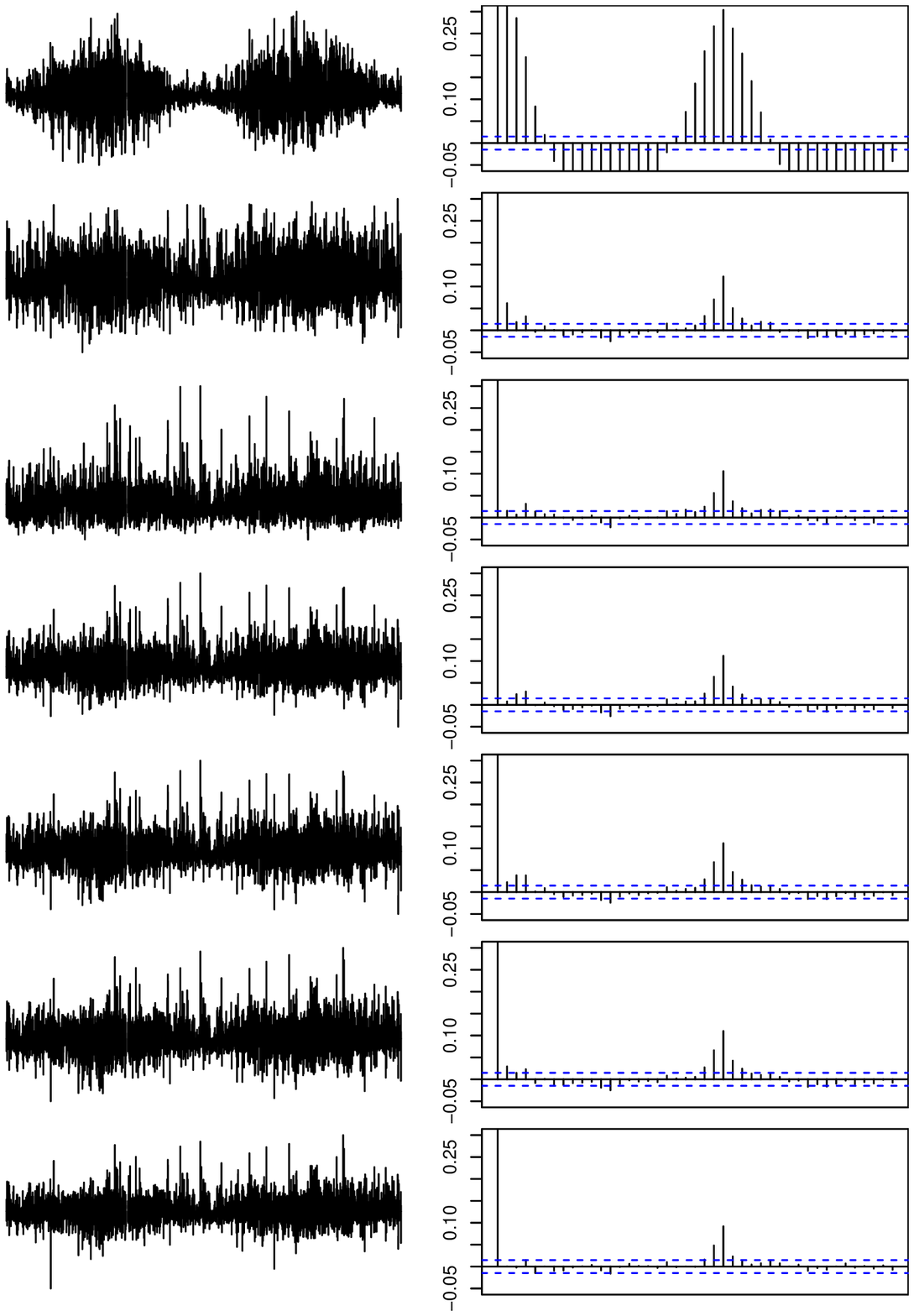}\\
    \vspace{-10pt}
  \caption{\emph{Studentized residuals and autocorrelation function for the residuals. Here the plots in the top line are from Model 1 continueing to the plots in the bottom line from Model 7.}}\label{fig:ACF}
\end{center}
\vspace{-10pt}
\end{figure}

An important analysis tool is the autocorrelation function. We compute this function for the studentized residuals. The autocorrelation for different lags for the different models is shown in Figure \ref{fig:ACF}. Observe that the range of the y-axis is  $(-0.05, 0.25)$ to better show the significance levels for the different lags. In Figure \ref{fig:ACF}, we see that Model 1 has many autocorrelation coefficients that are significant, indicating that this model clearly does not properly capture the dynamics of the solar irradiance process. In contrast, for Model 2, only the first lag is still significant. We then end up with model 7, in which the first 22 lags are insignificant in predicting the next time-step. Notice, however, that around lag number 24, we again begin to see significant autocorrelation coefficients. This is most likely caused by local conditions like shadowing (by trees or buildings) or local recurrent weather phenomena such as sea breeze (\cite{Bacher2013non-parametric}).

\begin{figure}[h!]
\begin{center}
  \includegraphics[width=0.8\textwidth]{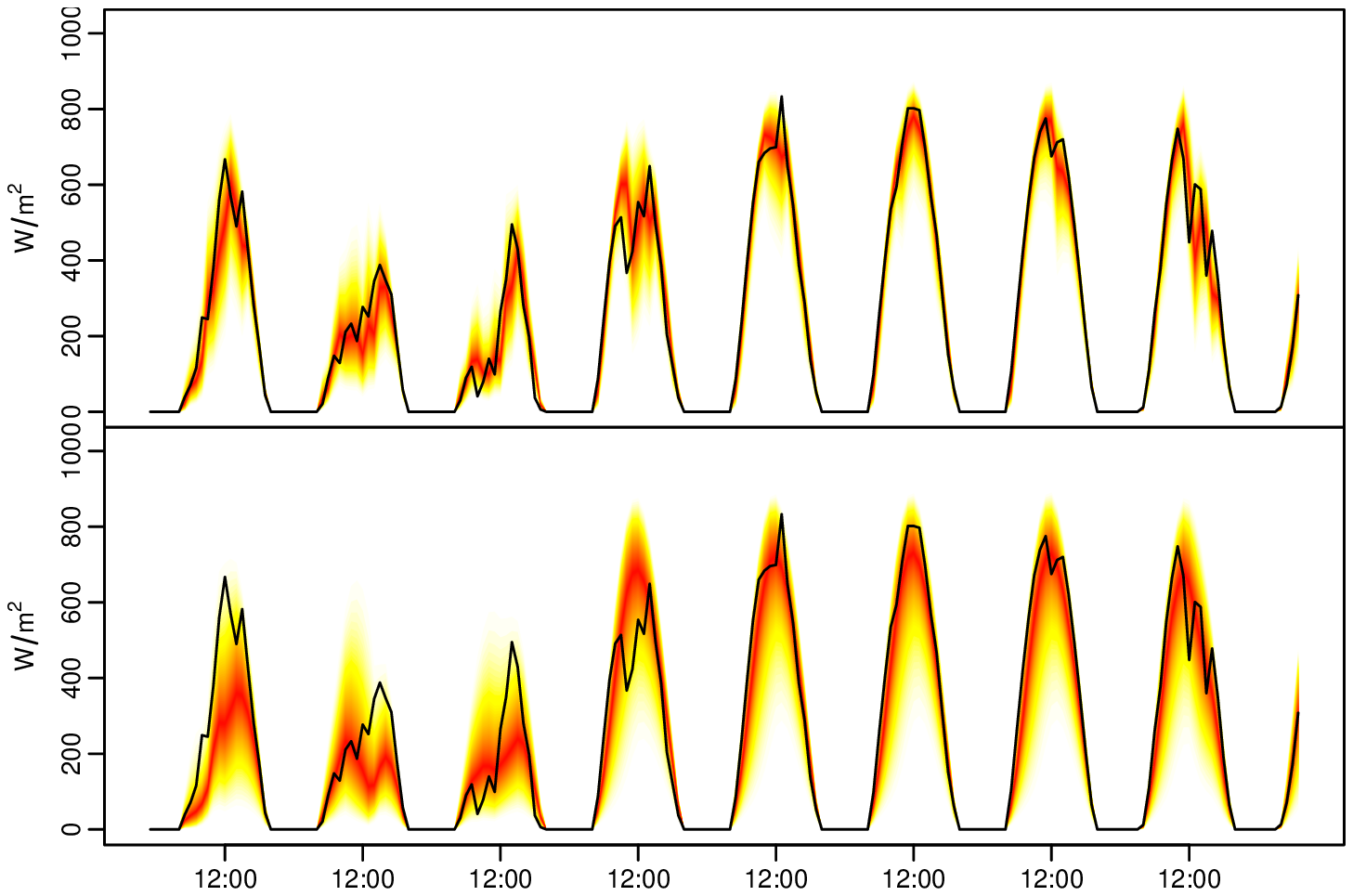}\\
    \vspace{-10pt}
  \caption{\emph{Observations of solar irradiance plotted together with the 1-h ahead predictive densities (top) and 24-h ahead (bottom). Warmer colors indicate a higher probability of seeing this realization.}}\label{fig:PredictiveDensity}
\end{center}
\vspace{-10pt}
\end{figure}
\begin{figure}[h!]
\begin{center}
  \includegraphics[width=0.8\textwidth]{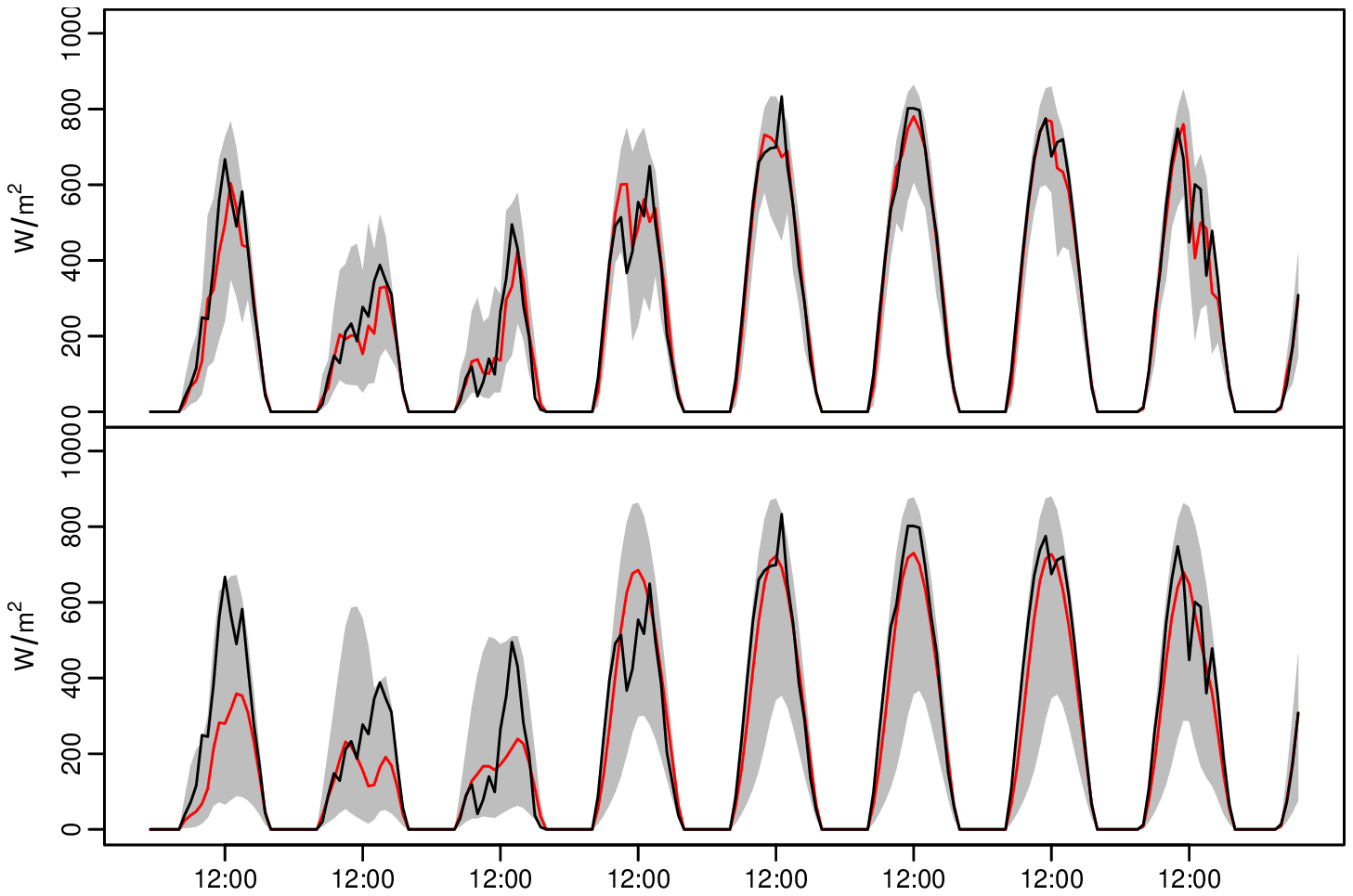}\\
    \vspace{-10pt}
  \caption{\emph{Observations of solar irradiance in black plotted along with the 1-h ahead (top) and 24-h ahead (bottom) 95 \% prediction intervals in gray and the prediction in red. }}\label{fig:95PredictionIntervals}
\end{center}
\vspace{-10pt}
\end{figure}

The output of the SDE models is the conditional predictive density at each point in time. In Figure \ref{fig:PredictiveDensity} the observations are shown along with the predictive densities given by Model 7, with warmer colors having higher probability. Firstly, we see that the conditional density seems to satisfactorily cover the observations and to be centered around them. Another feature of this model is that it assigns zero probability to events outside the state space, that is, for values of irradiance higher than the maximum or lower than zero. Notice, in the bottom plot of Figure \ref{fig:PredictiveDensity}, that the density spreads out, as here we represent predictions issued 24 hours ahead instead of 1 hour ahead. A further illustration of this is seen in Figure \ref{fig:95PredictionIntervals},
where the $95 \%$ prediction interval is shaded in gray. Note from this figure that this interval decreases as we approach the limits of the state space, that is, when the process comes closer to the maximum irradiance or to zero. Also, notice that the prediction interval is not symmetric around the point prediction, especially when approaching the limits. Upon careful inspection, it can be found that the prediction interval is the widest when we predict around $50\%$, which is to be expected from the physics of the system. Besides, notice that the 24-hour ahead forecast has a wider $95\%$ prediction interval.

To validate the accuracy of the predictive density, we can evaluate the predictive quantiles in the distribution. This is done by counting how many observations lie on each side of the predictive quantile in question and comparing it to the expected number.

\begin{table}[!ht]
\small
\begin{center}
    \begin{tabular}{rrrrrrrrr}
    \toprule
         & & \multicolumn{2}{c}{Training Set} & \multicolumn{2}{c}{Test Set}\\  \cmidrule(r){3-4} \cmidrule(r){5-6}
   Quantile function                     & Expected   & 1h 	    & 24h & 1h &  24h  \\ \midrule 
   $Q(0.1)$          & 0.10 & 0.088 & 0.079 & 0.076 & 0.061 \\
   $Q(0.2)$          & 0.20 & 0.176 & 0.170 & 0.156 & 0.141 \\
   $Q(0.3)$          & 0.30 & 0.273 & 0.261 & 0.253 & 0.220 \\
   $Q(0.4)$          & 0.40 & 0.376 & 0.348 & 0.349 & 0.301 \\
   $Q(0.5)$          & 0.50 & 0.486 & 0.440 & 0.458 & 0.392 \\
   $Q(0.6)$          & 0.60 & 0.603 & 0.540 & 0.589 & 0.473 \\
   $Q(0.7)$          & 0.70 & 0.720 & 0.645 & 0.712 & 0.580 \\
   $Q(0.8)$          & 0.80 & 0.818 & 0.763 & 0.811 & 0.728 \\
   $Q(0.9)$          & 0.90 & 0.901 & 0.885 & 0.902 & 0.858 \\
   \bottomrule
\end{tabular}
\end{center}
\vspace{-10pt}
  \caption{\emph{Frequency of observed exceedances for selected quantiles of the predictive density given by the quantile function $Q(\cdot)$.}}
\label{tab:Quantiles}
\end{table}

In Table \ref{tab:Quantiles} the exceedances of the predictive quantiles for Model 7 are shown. The predictive distribution is found by solving Equation (\ref{eg:Fokker-Plank}) for the estimated parameters and transforming the obtained density function to the observation space by the observation equation given by (\ref{eq: CTSM system 2}). In a perfect data fit, the expected quantiles match the observed ones exactly. For the 1-hour prediction on the training set, we see an excellent performance, with the frequency of exceedences quite close to the expected one in a perfect fit. For the 24-hour prediction horizon, we observe a slightly lower number of exceedances than expected. This is also true for the test set, especially for the 24-hour ahead quantiles. It should be taken into account that Model 7 has been fitted on the basis of one-step-ahead predictions, that is, the prediction of the next hour on the training set. Thus, the model is not tuned to predictions for a 24-hour horizon, even though it seems to perform reasonably well nevertheless.

\section{Concluding Remarks}

With the increasing penetration of renewable generation in energy systems, forecasting renewable production is becoming crucial for its efficient integration. Asymmetric costs in time associated with power generation further requires an understanding of the uncertainty associated with the renewable stochastic production. This paper proposes a stochastic differential equation framework for modeling the uncertainty associated with solar irradiance. It allows us to construct a process that is confined to a bounded state space and assigns zero probability to events outside this space, which is especially useful for probabilistic forecasting.

The starting point for the modeling done in this paper is a simple SDE that tracks the expected solar irradiance from a numerical weather prediction. By normalizing the weather prediction with the maximum irradiance, we can capture the periodic behaviour in the dynamics, and consequently, achieve major improvements. We can tune the diffusion term to model the actual behaviour of the process and confine it to a bounded interval. The SDE formulation allows for formulating complex model structures and to track conditional distributions at any point in time. Our proposed SDE modeling approach outperforms simple as well as more complex benchmarks.

Even though there is a relation between solar irradiance and produced power from a photovoltaic panel, such a relation is not trivial. It depends on tilt and orientation of the PV panel as well as on its efficiency, which may vary as the panel gets dirty or deteriorates over time. To address this, an adaptive estimation approach would be appropriate. Future studies will be directed at constructing models to capture this and produce probabilistic forecasts for solar power via a power curve. Since what is important to the energy system is the total input of renewable energy, future studies will also be directed at co-modeling wind and solar power, as these are expected to be main contributors to the energy mix of the future. Another potential line of future research is the modeling of the interdependence between the power output of solar farms at different locations. 

\newpage






%

\end{document}